\documentclass{article}

\usepackage[square,numbers]{natbib}

\usepackage[preprint]{macros/neurips_2020}

\usepackage[utf8]{inputenc} % allow utf-8 input
\usepackage[T1]{fontenc}    % use 8-bit T1 fonts
\usepackage{hyperref}       % hyperlinks
\usepackage{url}            % simple URL typesetting
\usepackage{booktabs}       % professional-quality tables
\usepackage{amsfonts}       % blackboard math symbols
\usepackage{nicefrac}       % compact symbols for 1/2, etc.
\usepackage{microtype}      % microtypography

\usepackage{amsmath}
\usepackage{graphicx}
\usepackage{dsfont}
\usepackage{dirtytalk}
\usepackage{bm}
\usepackage{import}
\usepackage{caption}
\usepackage{subcaption}

\hypersetup{colorlinks=true, linkcolor=blue, citecolor=blue, urlcolor = blue}

\def\reals{\mathbb{R}} % Real number symbol
\def\W{\bm{W}}
\def\w#1{\W^{(#1)}}
\def\y{\bm{Y}}
\def\Y#1{\y^{(#1)}}
\def\X{\bm{X}}
\def\K{\bm{K}}
\def\sv#1{\rho_{(#1)}^2}
\def\rv#1{\sigma_{#1}^2}
\def\q{Q}
\def\t{T}
\def\m{M}
\def\p{P}
\def\params{\bm{\theta}}
\def\spcov#1{\bm{\Psi}^{(#1)}}
\def\A{\bm{A}}

\def\ytrain#1{$\y^{\text{train }}_#1$}
\def\ytest#1{$\y^{\text{test }}_#1$}
\def\yseg{$\y^{\text{query }}$}
\def\xseg{$\X^{\text{query }}$}
\def\wtrain{$\W^{\text{train }}$}
\def\wtest{$\W^{\text{test }}$}
\def\xtrain#1{$\X^{\text{train }}_#1$}

\newcommand{\norm}[1]{\left\lVert#1\right\rVert}
\newcommand{\beginsupplement}{%
        \setcounter{table}{0}
        \renewcommand{\thetable}{S\arabic{table}}%
        \setcounter{figure}{0}
        \renewcommand{\thefigure}{S\arabic{figure}}%
        \setcounter{section}{0}
        \renewcommand{\thesection}{S\arabic{section}}
}
\author{%
  MohammadReza Ebrahimi $^{1,2}$\\
  \texttt{mr.ebrahimi@mail.utoronto.ca} \\
  \And
  Navona Calarco $^{2,4}$\\
  \texttt{navona.calarco@camh.ca} \\
  \AND
  Kieran Campbell $^{5}$\\
  \texttt{kieran.campbell@utoronto.ca} \\
  \And
  Colin Hawco $^{2,3}$\\
  \texttt{colin.hawco@camh.ca} \\
  \And
  Aristotle Voineskos $^{2,3,4}$\\
  \texttt{aristotle.voineskos@camh.ca} \\
  \And
  Ashish  Khisti $^{1}$\\
  \texttt{akhisti@ece.utoronto.ca} \\
}
\date{%
    $1$ Department of Electrical and Computer Engineering, University of Toronto, Toronto, ON, Canada\\
    $2$ Kimel Family Translational Imaging-Genetics Laboratory, Research Imaging Centre, \\
    \hspace{5pt} Centre for Addiction and Mental Health, Toronto, ON, Canada \\
    $3$ Department of Psychiatry, University of Toronto, Toronto, ON, Canada \\
    $4$ Institute of Medical Science, University of Toronto, Toronto, ON, Canada\\
    $5$ Department of Molecular Genetics, University of Toronto, Toronto, ON, Canada
}

\title{Time-Resolved fMRI Shared Response Model using Gaussian Process Factor Analysis}

\begin{document}
\maketitle
% \vspace{-1.5em}
\begin{abstract}
    Multi-subject fMRI studies are challenging due to the high variability of both brain anatomy and functional brain topographies across participants. An effective way of aggregating multi-subject fMRI data is to extract a shared representation that filters out unwanted variability among subjects. Some recent work has implemented probabilistic models to extract a shared representation in task fMRI. In the present work, we improve upon these models by incorporating temporal information in the common latent structures. We introduce a new model, Shared Gaussian Process Factor Analysis (S-GPFA), that discovers shared latent trajectories and subject-specific functional topographies, while modelling temporal correlation in fMRI data. We demonstrate the efficacy of our model in revealing ground truth latent structures using simulated data, and replicate experimental performance of time-segment matching and inter-subject similarity on the publicly available Raider and Sherlock datasets. We further test the utility of our model by analyzing its learned model parameters in the large multi-site SPINS dataset, on a social cognition task from participants with and without schizophrenia.
\end{abstract}

\section{Introduction}
\label{sec:intro}

Functional Magnetic Resonance Imaging (fMRI) is the most widely applied method to image human cognition. In it, neural hemodynamics are imaged in three dimensional space, over a fourth dimension of time. In task fMRI, participants additionally engage in an in-scanner experiment designed to engage some aspect of cognition. Of interest to neuroscientists are the observed topological and, increasingly, dynamical patterns associated with the given cognitive state. While fMRI has illuminated the `neural signatures' underlying many aspects of cognition, others, such as social cognition, remain elusive.

In recent years, neuroscientists have attempted to better capture true variability in human cognition. Larger and more representative imaging studies are increasingly the norm, boasting hundreds or even thousands of participants, often conducted across multiple research centres. These investigations have reported large individual differences in anatomical \cite{kanai2011structural} and functional networks, the latter in both the spatial \cite{mueller2013individual} and temporal domains \cite{davison2016individual}. Moreover, consensus evidence shows that task fMRI captures not only task-related cognition, but also background activity, physiological processes, and motion, as well as a variety of artifacts associated with scanner hardware \cite{liu2016noise}. In short: the observed fMRI signal represents a noisy amalgam of signals of varying and often unknown provenance.

Several Factor Analysis (FA) methods have been proposed to compensate for functional variability among subjects. For instance, Shared Response Model (SRM) \cite{srm} provides a means of aligning participants’ activation via a shared low dimensional response and subject-specific bases \cite{srm_enabling}, and Hierarchical Topographic Factor Analysis (HTFA) \cite{htfa} learns a global template of brain activity and casts each participant’s response as a perturbation of that template. However, these methods are limited in that they treat time as static, though converging evidence suggests that temporal correlation structure -- even on the slow timescale of hemodynamics -- carries meaningful signal \cite{infraslow}. Gaussian Process Factor Analysis (GPFA) \cite{gpfa} accommodates time by linking together factor analyzers in low-dimensional latent space, using imposed Gaussian process priors. This allows GPFA to model temporal and spatial structure in observation space, and has been applied within subjects to model dynamic functional connectivity \cite{lfgp}: an initial proof of principle of its suitability to fMRI.

% In this work, we introduce ‘Shared Gaussian Process Factor Analysis’ (S-GPFA) for time-resolved functional registration and subject aggregation. Our model exploits temporal dynamics to resolve the rotational ambiguity inherent in FA models. Thus, S-GPFA addresses the gaps detailed above: it anticipates and leverages high variability in large samples, it anticipates and parses noise via low-dimension representation, and crucially, it incorporates temporal structures in observed signals. We submit that S-GPFA may advance discovery of ‘neural signatures’ underlying human cognition. 

In this paper, we introduce ‘Shared Gaussian Process Factor Analysis’ (S-GPFA), a novel probabilistic model for  finding subject-specific topographies and shared temporal dynamics in multi-subject fMRI datasets. The proposed model simultaneously performs functional aggregation, dimensionality reduction, and dynamical modeling of fMRI data. We demonstrate the scientific utility of our model in identifying group-specific dynamical characteristics in the context of socioemotional cognitive capacity. Furthermore, identifying brain regions with meaningful functional variability across subjects within a heterogeneous sample provides preliminary evidence that S-GPFA can enable exploratory and hypothesis-driven examinations of functional topographies in psychiatric disorders.

\section{Background and Motivation}
\label{sec:related_models}
Shared Response Model (SRM) \cite{srm} is a multi-view extension of probabilistic Principal Component Analysis (pPCA) \cite{ppca}, wherein latent factors (named the shared response) are common across all subjects for each time point, while fixed factor loadings are specific to each subject. In addition, SRM explicitly imposes an orthogonality constraint on its loading matrices. Variants of this model such as Robust Shared Response Model (RSRM) \cite{rsrm} have been proposed in which both shared and private components in subjects' responses are explicitly modeled, similar to probabilistic Canonical Correlation Analysis (pCCA) \cite{pcca}. Similarly, in \cite{lukic2002ica} an ICA-based method has been proposed to separate shared and private sources in subjects' fMRI observations based on a similar factor model by exploiting time delayed correlations. 

Similar to static dimensionality reduction methods like Linear Factor Analysis (LFA) and PCA, SRM treats multivariate time series of fMRI recordings as a collection of independent snapshots in time, entirely disregarding temporal information. In other words, SRM remains invariant if training time series are shuffled through the time dimension. As with PCA, SRM cannot discover temporal dynamics unless such dynamics materialize as variance, which stems from both dynamics and noise \cite{dca}. 

Moreover, to address the rotational ambiguity of latent variables and factor loadings, SRM adopts a similar solution as PCA, by enforcing subjects' factor loadings to be orthonormal. Although this restricts the output to be unique, the resulting solution may not be interpretable since the assumption of orthogonal topographies is likely an undue simplification of physiology. Nonetheless, SRM's solution captures the shared structures in subjects' fMRI observations. An alternative solution for resolving rotational ambiguity is to utilize the inherent temporal dynamic of the data. This can address the aforementioned problems at once, i.e. finding unique and interpretable solutions whilst bringing insight about shared dynamic structures in observations. We adopt a similar approach as in Gaussian Process Factor Analysis (GPFA) \cite{gpfa}, where latent variables are linked through time by employing Gaussian Process priors, allowing viable modeling of both temporal and spatial covariance in data.
% Gaussian Process Factor Analysis (GPFA) [?] is an extension to FA that yields smooth low dimensional representation by utilizing time label information in data. More specifically, GPFA models latent factors as independent Gaussian Processes over time, which allows for the specification of a correlation structure across the low-dimensional states at different time points. [?] and [?] provide two recent application of GPFA in fMRI analysis.

\section{Shared Gaussian Process Factor Analysis Model (S-GPFA)}
\label{sec:sgpfa}

\subsection{Model description}
\label{subsec:model}
\begin{figure}[t]
    \begin{minipage}{.5\textwidth}
        \centering
        \includegraphics[width=\linewidth]{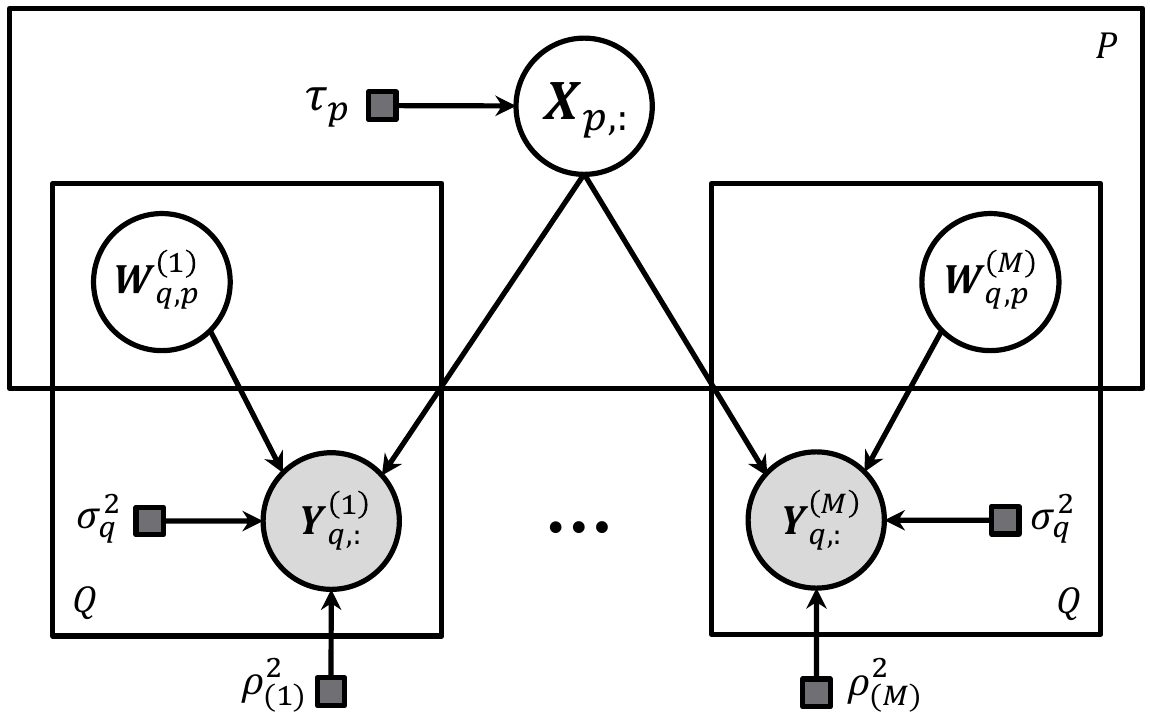}
        \caption{Graphical model for S-GPFA.} \label{fig:model}
    \end{minipage}
    \begin{minipage}[c]{.5\textwidth}
        \savebox\strutbox{$\vphantom{\dfrac11}$}
        \begin{align*}
            & \X_{p, :} \sim \mathcal{GP}\left(0, \kappa_p(.,.) \right) \\
            & \w{m}_{q,p} \sim \mathcal{N}(0, 1) \\
            & \Y{m}_{:, t} | \X_{:, t}, \w{m} \sim \mathcal{N} \left( \w{m} \X_{:, t}, \spcov{m} \right) \\
            & \kappa_p(t_1, t_2) = \exp \left( -\frac{(t_1 - t_2)^2}{2\tau_p^2} \right)
        \end{align*}
  \end{minipage}
\end{figure}

In the following, we present the mathematical formulation of the probabilistic model for S-GPFA. We denote by $\A_{i,:}$, $\A_{:, j}$, and $[\A]_{m,n}$ the $i$th row vector, the $j$th column vector, and element $(m, n)$ of matrix $\A$, respectively. Let $\Y{m} \in \reals^{\q \times \t}$ be the observed high-dimensional fMRI time series for subject $m \in \{1, 2, \ldots, \m \}$ where $\q$ is the number of voxels (or in general, brain regions), $\t$ is the number of time samples (in TRs), and $\m$ is the total number of subjects in the dataset. We denote by $\Y{m}_{q,:} \in \reals^{1\times \t}$ the $q$-th row of the observation matrix, or equivalently, the time series of brain activation in region $q$ for subject $m$. We assume all subjects are exposed to identical and time-synchronized stimuli while brain activities are recorded and activation time series are centered over time.
S-GPFA extracts shared low-dimensional latent trajectories $\X \in \reals^{\p \times \t}$ describing the common latent state of all subjects, where their linear combinations through loading matrices of each subject describe the observed activation time series.
Here, $\p$ is a hyperparameter to be set as the dimensionality of the latent space ($\p < \q$). 
Following the standard Factor Analysis model, we define a set of linear (isotropic) Gaussian systems for fMRI observations and latent trajectories:
\begin{equation}
    % \y_{q, :} | \X \sim \mathcal{N} ( \W_{q, :}^{(m)}\X, \rho_{m}^2 \sigma_q^2 I)
    \Y{m}_{:, t} | \X_{:, t}\, , \w{m} \sim \mathcal{N} \left( \w{m} \X_{:, t}, \spcov{m} \right)
\end{equation}
where $\w{m} \in \reals^{\q \times \p}$ and $\spcov{m} \in \reals^{\q \times \q}$ denote the factor loading matrix and observation noise covariance matrix for subject $m$, respectively. Note that columns of factor loadings can be considered as brain activation bases meant to capture subject-specific topographies (we use the terms `factor loadings' and `subject topographies' interchangeably). Constraining $\spcov{m}$ to be diagonal will let $\w{m}$ and $\X$ completely define the covariance structure of region activation patterns.
% The $q$th diagonal element in $\spcov{m}$ represents the independent noise variance of region $q$ in subject $m$.
To explicitly let brain region $q$ of subject $m$ accommodate different noise levels, we model the diagonal elements as $[\spcov{m}]_{q, q} = \sv{m} \rv{q}$, where $\rho$ and $\sigma$ are subject-specific and region-specific learnable parameters, respectively. Following standard GPFA, to model the temporal correlation of the data, we impose independent Gaussian Process (GP) priors over each latent trajectory:
\begin{equation}
    \X_{p, :} \sim \mathcal{GP}\left(0, \kappa_p(\cdot , \cdot) \right), \hspace{1.5em} p \in \{1, \ldots, \p\}
\end{equation}
where $\kappa_p$ denotes a Mercer kernel function. Therefore, the covariance matrix for the $p$th latent trajectory, $\K_p$, will be the Gram matrix of the kernel over the index set $\{1, 2, \ldots, \t\}$, i.e. $[\K_p]_{t_1,t_2} = \kappa_p(t_1, t_2)$. In general, the choice of kernel function imposes important assumptions on the form and smoothness of observed time series. Following prior literature \cite{infraslow,lfgp}, we opt to employ commonly-used Squared Exponential (SE) kernel functions:
\begin{equation}
    \kappa_p (t_1, t_2) = \alpha_p^2 \exp \left( -\frac{(t_1 - t_2)^2}{2\tau_p^2} \right) + \eta_p^2 \, \mathds{1}_{t_1=t_2}
\end{equation}
Hence, the GP prior is fully parameterized through the kernel variance $\alpha_p^2$, characteristic timescale $\tau_p \in \reals_+$, and the kernel independent noise variance $\eta_p^2$. Furthermore, we fix the scale of $\X$ to have $\X_{:,t} \sim \mathcal{N}(0, I)$ by setting $\alpha_p^2 + \eta_p^2 = 1$. This will prevent the identifiability issue in the scale of $\X$ and $\w{m}$ by allowing unconstrained learning for factor loadings \cite{gpfa,infraslow,lfgp}. We also set $\eta_p^2$ to a small value to act as a diagonal jitter for numerical stability. Therefore, the characteristic timescales $\tau_p$ can fully define the priors over latent trajectories. Figure \ref{fig:model} shows the graphical model for S-GPFA along with the summary of model specification.
% S-GPFA models within and between-subjects temporal and spatial covariance in a fMRI dataset: 
% \begin{equation}
%     \operatorname{Cov} \left( \Y{m_1}_{q_1, t_1} , \Y{m_2}_{q_2, t_2} \right) = \sum_{p=1}^{\p} \w{m_1}_{q_1,p} \w{m_2}_{q_2,p} \kappa_p(t_1, t_2) + \delta_{m_1, m_2} \delta_{q_1, q_2} \delta_{t_1, t_2} \rv{q_1} \sv{m_1}
%     % \operatorname{Cov} \left( \Y{m}_{q, t} , \Y{m'}_{q', t'} \right) = \w{m}_{q,:} [] \w{m'}_{q',:}^T + \delta_{m, m'} \delta_{q, q'} \delta_{t, t'} \rv{q} \sv{m}
% \end{equation}

To find the model parameters, we employ gradient ascent (using ADAM optimizer \cite{adam} and TensorFlow Probability) to maximize the joint probability distribution of observations and latent variables, conditioned on model parameters. The collective set of model parameters and latent variables to be inferred is denoted by $\params=\left\{ \X, \{\w{m}, \sv{m}\}_{m=1}^M, \{\rv{q}\}_{q=1}^Q, \{\tau_p\}_{p=1}^P \right\}$. The resulting maximum a posteriori (MAP) estimate of model parameters is in the form of equation (\ref{eq:nll}):
\begin{align}
    \hat\params = \underset{\params}{\operatorname{argmin}} & \sum_{m=1}^{M} \sum_{q=1}^{Q} \left[ \frac{T}{2} \log(2\pi\sv{m}\rv{q}) + \frac{\norm{ \Y{m}_{q,:} - \w{m}_{q,:} \X }^2}{2\sv{m}\rv{q}} \right]  \nonumber \\
    + & \lambda \sum_{p=1}^{P} \left[ \frac{1}{2}\log\det(2\pi \K_{p}) + \frac{1}{2}\X_{p,:} \K_{p}^{-1} \X_{p,:}^T \right] + \sum_{m=1}^{M} \frac{1}{2} \norm{\w{m}}_F^2 \label{eq:nll}
\end{align}

where $\norm{\cdot}_F$ denotes the Frobenius norm. For now, let us set $\lambda=1$ to achieve the standard MAP solution. We will explain the motivation behind including such constant in the objective function in section \ref{sec:lambdaloss}. The first summation term in (\ref{eq:nll}) is the reconstruction loss promoting data fit. We refer to the second summation term as the smoothness loss, where $\log\det(\K_p)$ constitutes a model complexity penalty (in terms of smoothness) and $\X_{p,:} \K_{p}^{-1} \X_{p,:}^T$ promotes smooth latent trajectories governed by $\tau_p$. Finally, the last term is a weight decay loss for subjects' factor loadings. While the reconstruction loss favors complex models to fit the observed data, smoothness and loadings weight decay losses act as regularizers to promote smooth and less flexible models, respectively.

Adding a new subject $\m+1$ to a previously trained model is done through finding the maximizer of $\text{P}(\Y{\m+1}, \w{\m+1} | \X, \sv{\m+1},  \{\rv{q}\}_{q=1}^Q)$ with respect to $\w{\m+1}$ and $\sv{\m+1}$. Note that the shared latent trajectories, existing subject's topographies, and region/subject noise factors will remain unchanged.
In addition, using previously learned topographies, shared timescales, and noise factors, one can map new observations from the existing subjects in the training cohort into the shared space. This can be done by maximizing $\text{P}(\{\y_{\text{new}}^{(m)}\}_{m=1}^{\m}, \X_{\text{new}} | \{\w{m}, \sv{m}\}_{m=1}^M, \{\rv{q}\}_{q=1}^Q, \{\tau_p\}_{p=1}^P)$ with respect to $\X_{\text{new}}$, where $\y_{\text{new}}$ denotes new observations from the subjects in the training cohort and $\X_{\text{new}}$ shows the associated shared latent trajectories. Similarly, note that previously-trained subject's topographies, shared latent timescales, and region/subject noise factors will remain unchanged.
% \vspace{-5pt}
\subsection{Modified training objective}
\label{sec:lambdaloss}

\begin{figure}[tb!]
    \centering
    \includegraphics[width=\linewidth]{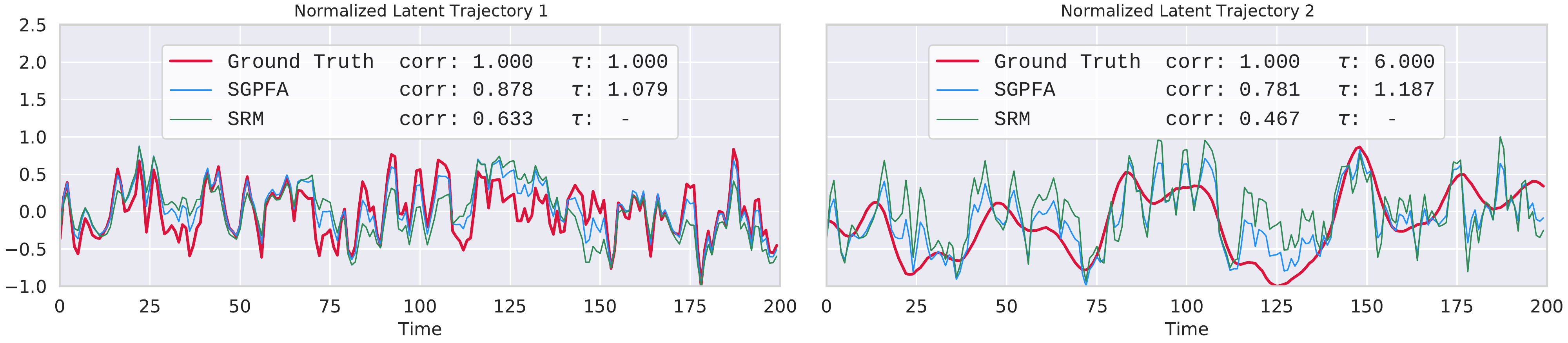}
    \includegraphics[width=\linewidth]{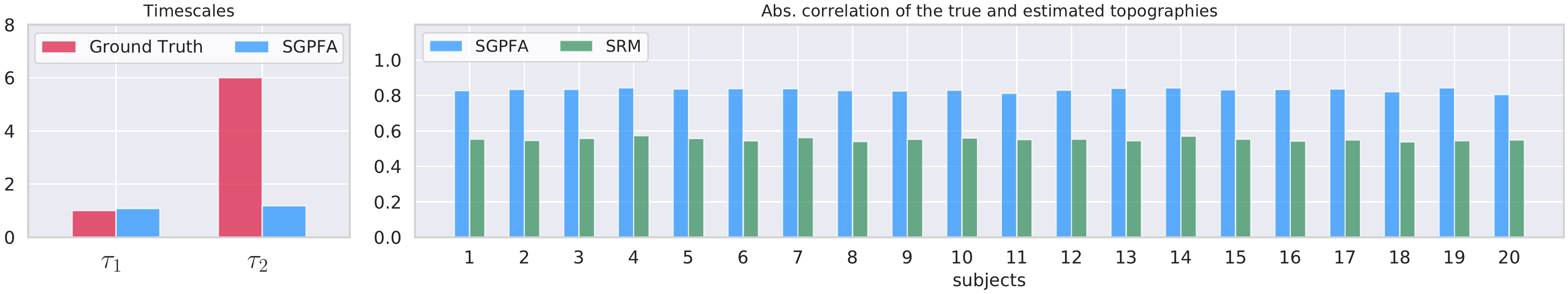}
    \vspace{-1.5em}
    \caption{$\lambda=1$: Simulated dataset with parameters $M=20$, $Q=50$, $T=200$, $P=2$. Latent trajectories sampled from independent GP with SE kernel and fixed timescales $\tau_1=1, \tau_2=6$. Factor loadings, as well as subject and region noise levels are sampled from a standard normal distribution.} \label{fig:sim_sgpfa_1}
    \vspace{15pt}
    \includegraphics[width=\linewidth]{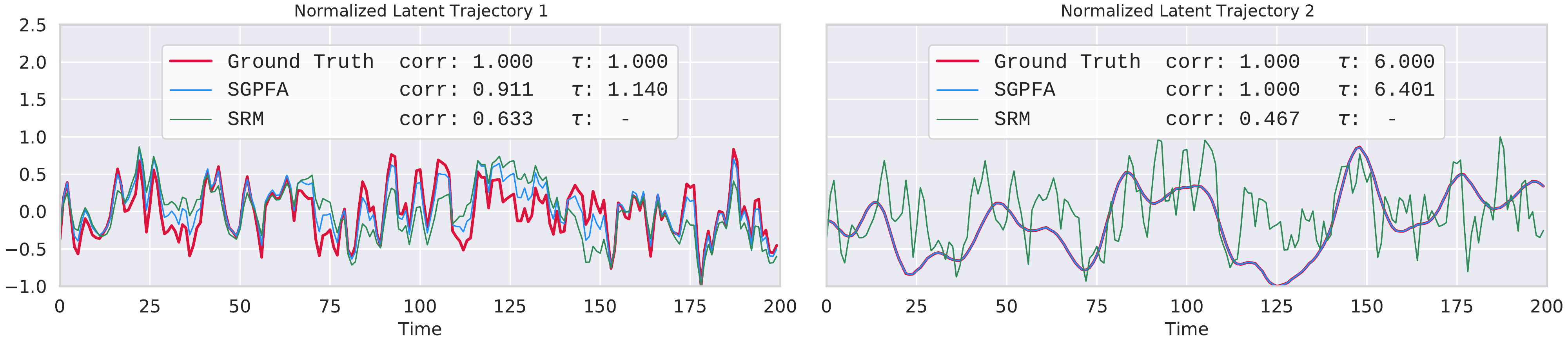}
    \includegraphics[width=\linewidth]{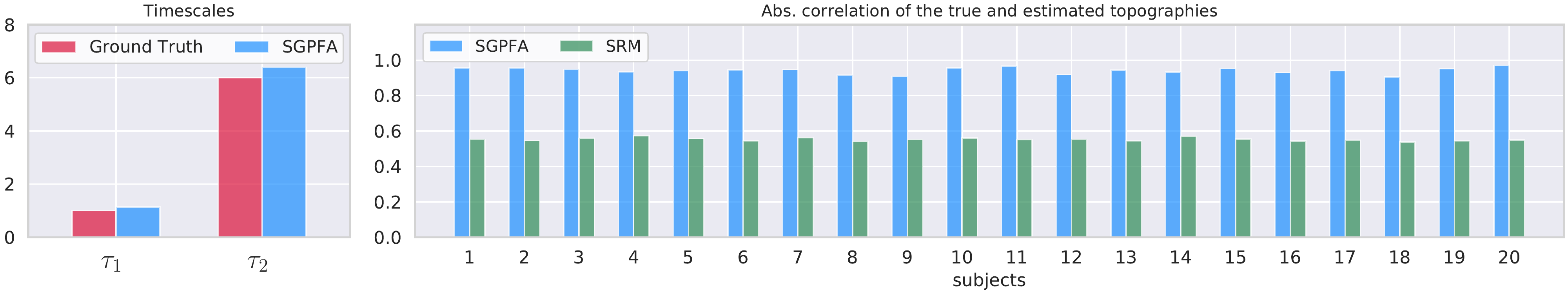}
    \vspace{-1.5em}
    \caption{$\lambda=0.1 \times \m\q/\p=50$. Same dataset as in Figure \ref{fig:sim_sgpfa_1}. Amplified smoothness loss results in finding more accurate parameters.} \label{fig:sim_sgpfa_2}
    % \vspace{-5pt}
\end{figure}

Increasingly, fMRI studies include sample sizes of hundreds of participants ($M$) with thousands of voxels recorded ($Q$). An examination of the objective function in (\ref{eq:nll}) reveals that the objective is extensively dominated by the reconstruction loss, since the total number of observed time series, $Q M$, is often orders of magnitudes larger than the latent space dimensionality $P$. In such a case, the solution of (\ref{eq:nll}) cannot afford to model the temporal smoothness properties of the data, and consequently, attempts to fit the observed data with a possibly complex model. To resolve this issue, we amplify the smoothness loss in (\ref{eq:nll}) by hyperparameter $\lambda \propto \m\q/\p$ to balance the weight of the smoothness loss against the reconstruction loss. Note that from a probabilistic perspective, such regularization can be interpreted as weighted likelihood \cite{weightedlikelihood}. Although, as with any hyperparameter, standard methods like cross validation can be employed to tune $\lambda$, we use the fixed value of $0.1 \times \m\q/\p$ for all experiments conducted in the present paper. 

Using simulated data, we can demonstrate how amplifying the smoothness loss helps the model use temporal dynamics of the observations to accurately estimate shared latent responses and subject-specific topographies. But prior to this, we should address two inherent identifiability issues. First, the order of latent variables is naturally arbitrary, and second, the signs of shared latent trajectories $\{\X_{p,:}\}_{p=1}^{\p}$ and subject topographies $\{\W_{:,p}\}_{p=1}^{\p}$ are in opposite interplay. Moreover, in SRM, factor loadings are constrained to be orthonormal, hindering direct comparison of the true and estimated parameters. However, neither the order nor the scale and sign of these parameters are of interest. Therefore, to deal with the former issue, for each model we choose the ordering of latent dimensions that maximizes the sum of absolute correlations between matched pairs of the true and estimated latent time series. To address the latter issue, we use the absolute value of correlations between the ground truth and estimations to assess the performance of each model. 

To generate simulated data, we sample $\p = 2$ latent trajectories from independent Gaussian Processes (with SE kernel and fixed timescales). We also sample factor loading weights from a standard normal distribution, independently for each of $M=20$ subjects. Linear combination of latent trajectories and factor loadings generate $Q = 50$ dimensional times series of length $T = 200$ for each subject. Noisy versions of these observations are used to train S-GPFA and SRM models. We use the optimized implementation of SRM \cite{srm_enabling} from the Brain Imaging Analysis Kit.
Figures \ref{fig:sim_sgpfa_1} and \ref{fig:sim_sgpfa_2} show ground truth versus estimated shared latent trajectories and timescales, as well as the correlation between the true and estimated subject topographies for S-GPFA, with and without smoothness amplification. As Figure \ref{fig:sim_sgpfa_1} depicts, with $\lambda=1$, although S-GPFA outperforms SRM in terms of revealing true latent trajectories and subject topographies, it is unable to accurately discover the temporal dynamic of the shared responses, rendering the estimated latent trajectories and subject topographies non-informative. However, as results shown in Figure \ref{fig:sim_sgpfa_2} suggest, increasing $\lambda$ not only allows S-GPFA to estimate timescales and shared latent trajectories accurately, but it also increases the accuracy of determining subject topographies, resulting in more meaningful -- and in practice, more interpretable -- description of observed data. More examples and comparisons of model performance using simulated data are presented in Supplementary Materials.

\section{Experiments}
\label{sec:experiments}

We conducted three experiments to demonstrate the utility of S-GPFA. First, inter-subject similarity, allows us to test the ability of our model in finding dynamical components shared among all subjects' recordings. Second, time segment matching, measures generalization of learned temporal dynamics to unseen subjects and observations. In the third experiment, we apply our model to a multi-subject dataset, allowing us to evaluate and interpret subjects' topographies and group-specific temporal dynamics.
% Taken together, our experiments show the proposed model can achieve equal performance in terms of aggregating subjects' observations, by employing temporal information already available in data, without the need for imposing unrealistic modeling assumptions. 

The first two experiments use the Raider \cite{ha} and Sherlock \cite{sherlock} datasets. The Raider dataset collects recordings of 1000 voxels from the ventral temporal cortex, for 10 healthy adult participants passively watching the full-length movie ``Raiders of the Lost Ark''. In the Sherlock dataset, 17 subjects watched an episode of the TV series ``Sherlock'' while 481 voxels from posterior medial cortex were recorded. The third experiment uses the SPINS dataset, which records whole brain activity, for 332 adults with (187 subjects) and without (145 subjects) schizophrenia, watching video vignettes of emotional narratives, while providing real-time ratings of emotional valence. Relevant fMRI acquisition and preprocessing steps are presented in the Supplementary Material;  further details of both datasets, including inclusion criteria and demographic and clinical characterization, has been published elsewhere by original study authors \cite{ha, sherlock, spins}.

\subsection{Inter-subject similarity}
% The message of this experiment is, S-GPFA attempts finding shared "dynamical patterns of activations" [?] among subjects, while SRM finds a common embedding for participants data. It is expected to find less similarity. 

As suggested in \cite{srm}, we may examine the extent to which a shared response exists among subjects in observation space as follows: Leaving one subject out, we find the average response of the remaining subjects; next, we calculate the Pearson correlation over voxels between the left-out and averaged responses for each timepoint. Averaging over all combinations of left-out subjects (in Fisher Z domain) provides a measure of inter-subject similarity for every time point.

\begin{figure}[t!]
    \centering
    \includegraphics[width=\linewidth]{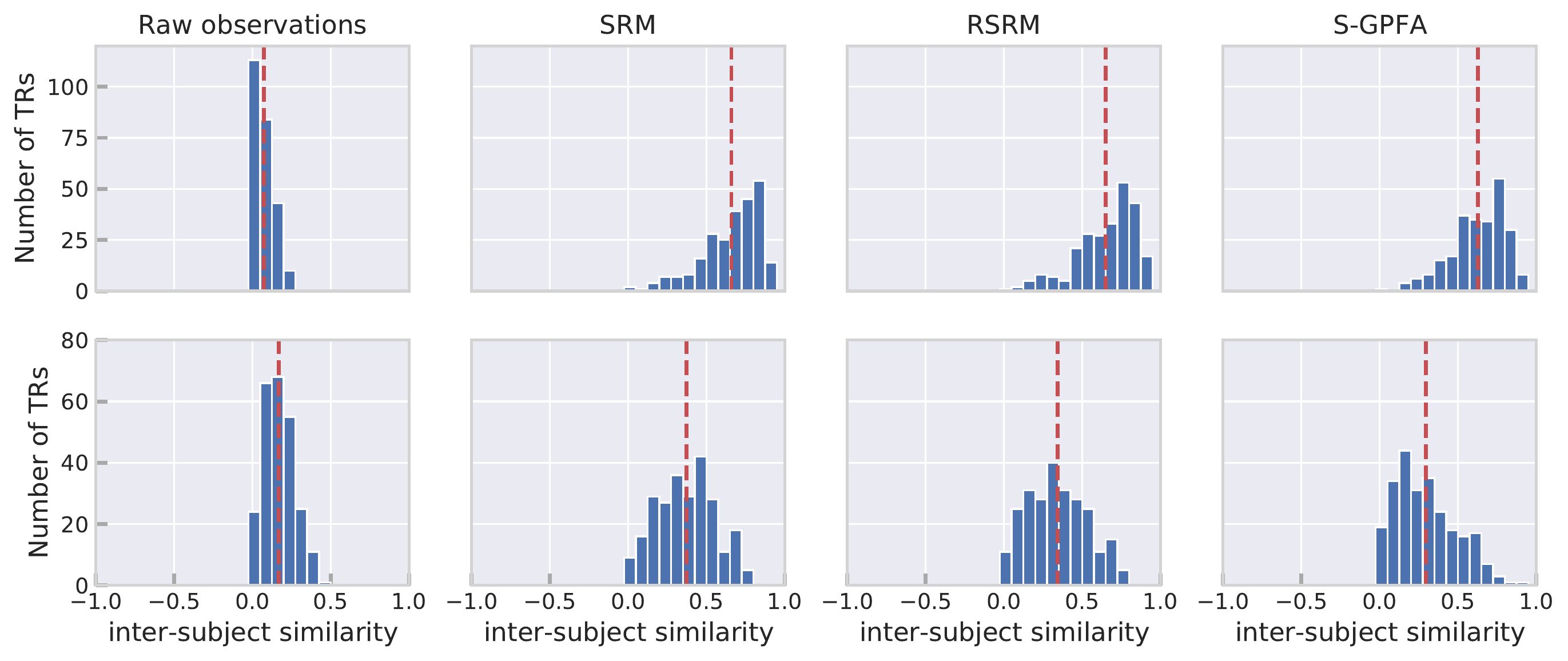}
    % \vspace{-0.5em}
    \caption{Inter-subject similarity. Top: Raider dataset (10 subjects, first 500 TRs). Bottom: Sherlock dataset (17 subjects, first 500 TRs), comparing raw observations, SRM, RSRM, and S-GPFA. We used $\p=10$ shared space dimensions. Dashed line indicates mean similarity over TRs.}\label{fig:isc}
    % \vspace{-5pt}
\end{figure}

In order to examine the ability of our model to find shared temporal components of participants' fMRI recordings, we calculated inter-subject similarities in the shared space learned by the model. In this case, data from all subjects were split in time into two equally-sized parts. Using the first part, we trained a model to learn subject-specific topographies. Next, we mapped the second part of the data into the shared space using the learned topographies, separately for each subject (section \ref{subsec:model}). Finally, to find the similarity of mapped responses at each time point, we performed a similar leave-subject-out procedure in the shared space by finding Pearson correlations of each subject and the average of the other subjects, over latent space components. Figure \ref{fig:isc} shows histogram plots of inter-subject similarities for the Raider dataset, calculated in voxel space as well as shared spaces learned by S-GPFA and SRM. Observed similarity in the shared embedding space found by S-GPFA confirms that temporal dynamics found by our model are common over all subjects.

\subsection{Time segment matching}
A time segment matching experiment, as first introduced in \cite{ha}, can evaluate how shared dynamics found by S-GPFA generalizes to new subjects and unseen data. We follow the modified version of the experiment presented in \cite{srm}, where the task is to locate an unseen segment of a test subject's response in time. We partition the fMRI dataset in time into two equal-size parts and leave one subject out to use the remaining subjects as training participants. We shall note the four resulting parts as \ytrain{1}, \ytest{1}, \ytrain{2}, and \ytest{2}.

\begin{figure}[tb!]
    \centering
    \begin{subfigure}[b]{0.45\textwidth}
        \centering
        \includegraphics[width=\textwidth]{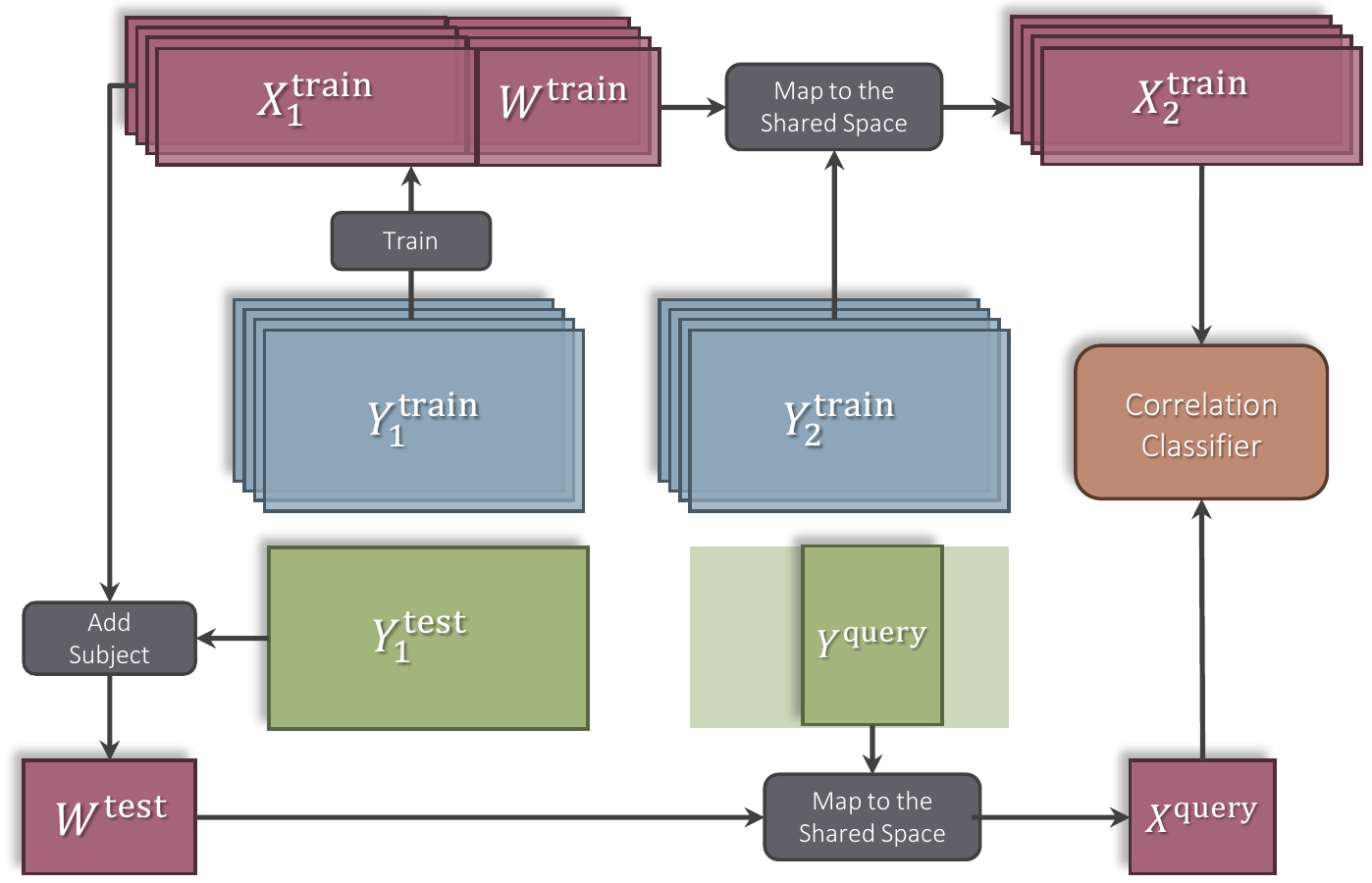}
        % \caption{}
        \label{fig:tsm_block}
    \end{subfigure}
    \hfill
    \begin{subfigure}[b]{0.54\textwidth}
        \centering
        \includegraphics[width=\textwidth]{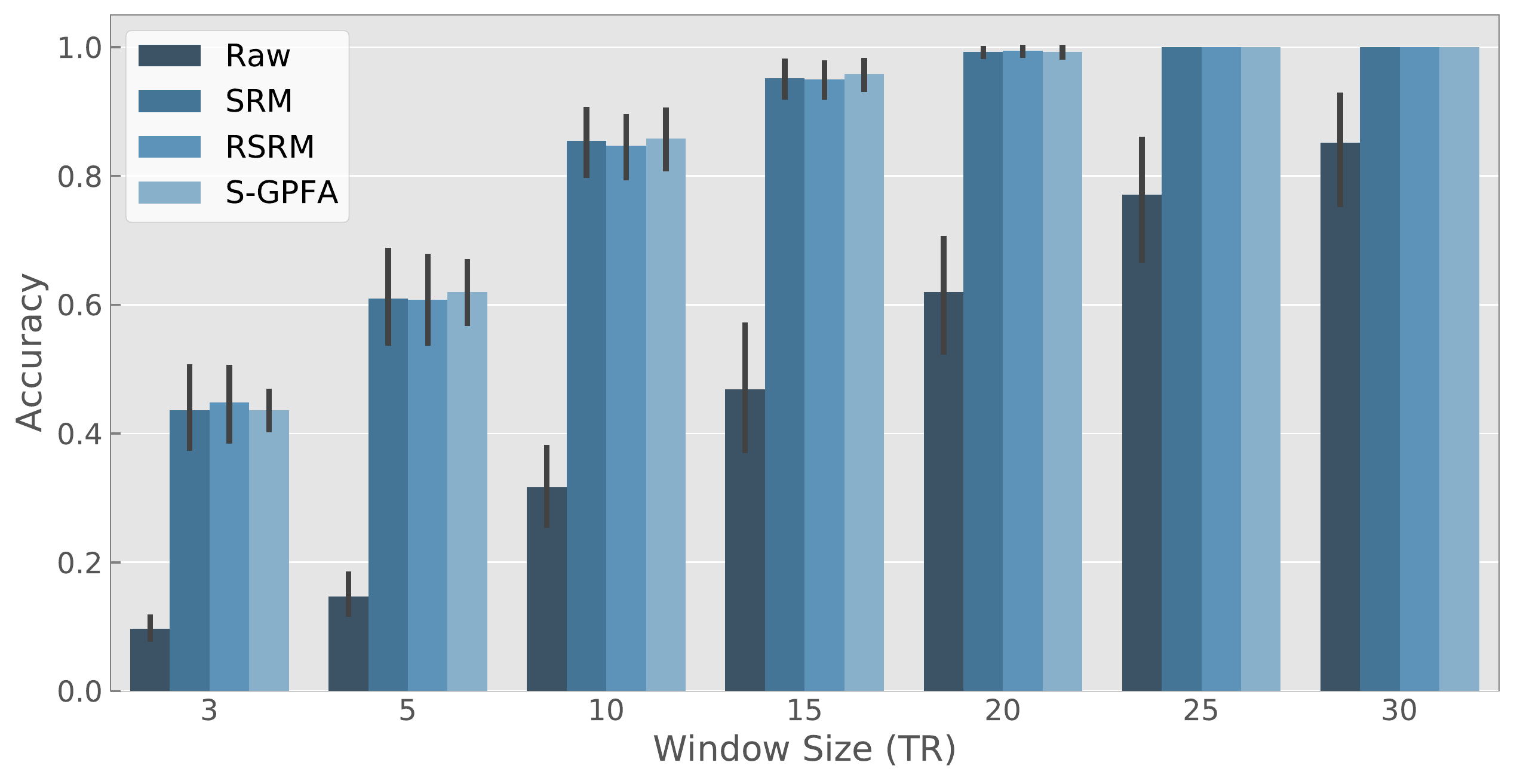}
        % \caption{}
        \label{fig:tsm_bar}
    \end{subfigure}
    \hfill
    \vspace{-15pt}
    \caption{Time segment matching experiment. Left: schematic procedure of the experiment. Right: Time segment matching accuracy for Raider dataset (10 subjects, first 400 TRs). We used $P= 10$ for SRM \cite{srm}, RSRM \cite{rsrm}, and S-GPFA. Error bars show $\pm$ standard deviations.}
    \label{fig:tsm}
\end{figure}

As shown in the left diagram in Figure \ref{fig:tsm}, the first half of the data is used to learn subject-specific topographies. First, an S-GPFA model is fit to the first half of training subjects' data \ytrain{1}, to learn shared latent trajectories \xtrain{1} and subject-specific topographies \wtrain. Next, using the learned shared trajectories \xtrain{1} and the first half of test subject's data \ytest{1}, topographies for the held-out subjects \wtest is calculated, as discussed in section \ref{subsec:model}. The second half of data is used for measuring time segment matching accuracy. More specifically, given the second half of training subjects' data, \ytrain{2}, we wish to locate an unknown segment from the held-out subject's data \yseg in time. By only using the raw observations as a baseline, one can first find the average response of training subjects, then report the time where \yseg and the average response are maximally correlated. Moreover, using learned subject topographies, we can transform \ytrain{2} and \yseg into the shared space using \wtrain and \wtest as discussed in section \ref{subsec:model}, to find \xtrain{2} and \xseg. Using a similar correlation classifier, we can locate the test segment in the point where \xtrain{2} and \xseg are maximally correlated. Figure \ref{fig:tsm} compares the time segment matching accuracy for different query segment lengths. S-GPFA demonstrates similar performance as SRM in terms of time segment matching accuracy. This posits that shared temporal dynamics (timescales) found in training subjects can be generalized to new subjects with unseen observations. 

\subsection{Application: SPINS dataset}
\paragraph{Dataset description}
For our final set of experiments, we applied S-GPFA to a subset (332 subjects) from the NIMH Data Archive study “Social Processes Initiative in the Neurobiology of the Schizophrenia(s)” (SPINS) who completed the Emotional Accuracy (EA) task. The EA task collects fMRI as participants watch videos of an actor (‘target’) recounting autobiographical events. In total, 9 videos were shown, lasting for between 2-2.5 minutes. Participants provide ratings of the target’s valence on a 9-point scale (1=extremely negative, 9=extremely positive) in real time via button press. The tasks' primary dependent measure, the EA score, is the correlation between the participant’s ratings of the targets’ emotions, and the “gold standard” rating of the targets’ ratings of their own emotions, calculated in 2 second time epochs. We selected SPINS as an experimental dataset because it includes participants with and without schizophrenia, and we held that the anticipated variability in brain structure, function, and cognitive performance would provide an interesting test of S-GPFA.

% \vspace{-5pt}
\subsubsection{Consistency of functional topographies}
\label{subsec:topo_consistency}
\begin{figure}[t]
    \centering
    \includegraphics[width=1\linewidth]{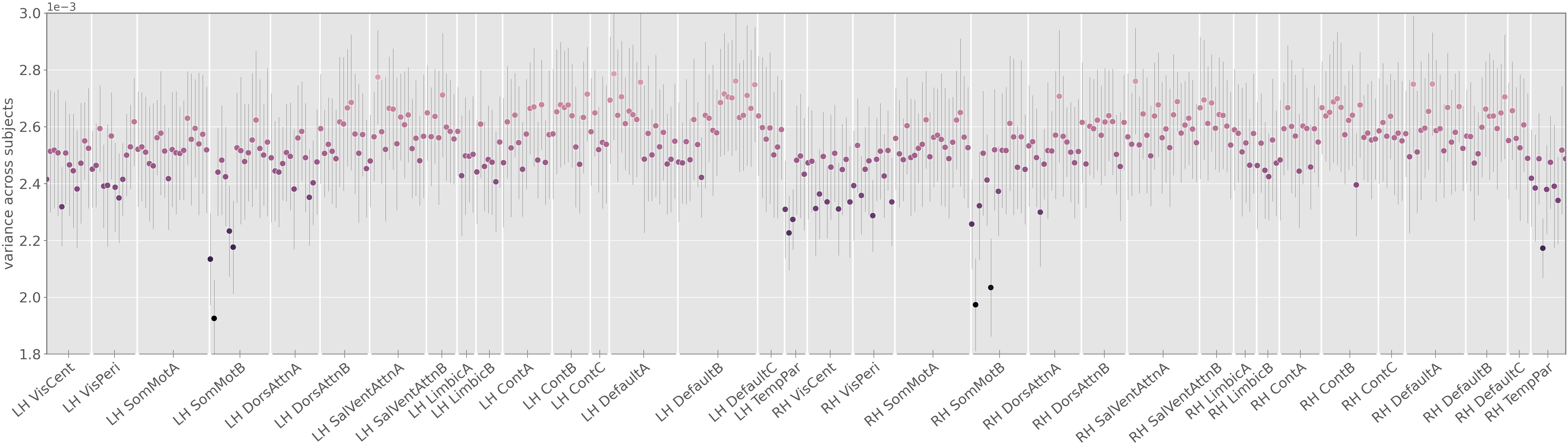}
    \includegraphics[width=1\linewidth]{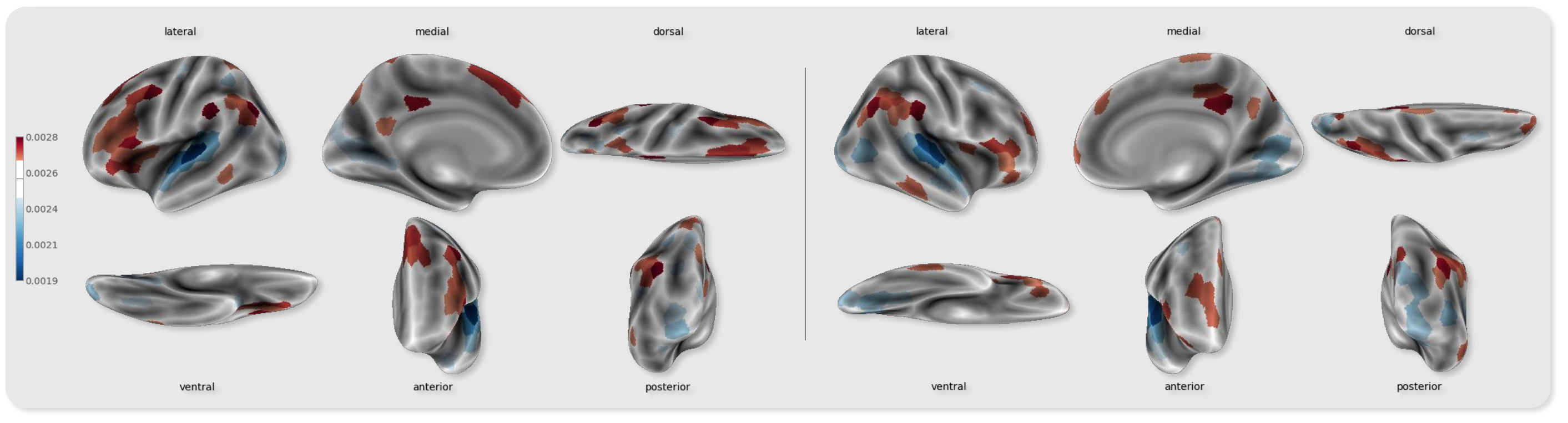}
    % \vspace{-0.5em}
    \caption{Top: Variance of normalized $\p=20$ topographies across subjects (Y axis) for different regions (X axis). Regions are divided into 17 Yeo sub-networks \cite{yeo} (See Fig. \ref{fig_supmat:schaefer}). Colored dots indicate the mean (across different topographies) of variances for each region, and error bars shows $\pm$ standard deviation. Bottom: Brain surface plots of average variance of normalized topographies, for left and right hemispheres. Blue (red) indicates higher (lower) consistency across subjects. In both panels, the second of nine EA videos is shown: all videos showed comparable patterns. See section \ref{sec_suppmat:topo_consistency} for other EA videos.}\label{fig:topo}
    % \vspace{-7pt}
\end{figure}

In our first analysis, we examined whole-brain variation in activation during the EA task, across all subjects. Learned factor loading matrices $\W$ in S-GPFA are, by model definition, subject-specific basis vectors that generate each subject's observation from a shared set of latent trajectories: $\w{m}\X \simeq \Y{m}$. Therefore, columns of factor loadings $\{\W_{:,p} \in \reals^\q\}_{p=1}^{\p}$, for different subjects, are perturbed versions of an activation template. This allows the model to capture functional variability across subjects. Hence, we can examine the amount of subject variability in each individual brain region of interest (ROI) for a given task, by looking at the variance of learned topographies over participants. To this end, for every topography $p$ of subject $m$, we first normalize $\w{m}_{:,p}$, then we find the variance of these normalized topographies over subjects. This will result in $\p$ variance values for each of $\q$ regions indicating a measure of subject variability, as shown in Figure  \ref{fig:topo}. Note that, for the present analysis, we used cortical parcellations from the Schaefer atlas (400 ROIs) \cite{schaefer2018local}, grouped in accordance with the Yeo 17 network parcellation \cite{yeo}. 

Examination of subject variance in functional topography itself shows variability across ROIs. Perhaps unsurprisingly, the lowest variance (blue) is evident in the motor network, engaged similarly by the task’s continuous button-press demands, and in temporal lobe areas related to audition, engaged by the audio aspect of the task. Excitingly, we see lower consistency (red) in the distributed frontal-parietal (‘Control A’) network, which contains the inferior frontal gyrus (IFG) and inferior parietal lobule (IPL), that together constitute the canonical human mirror neuron system \cite{iacoboni2006mirror}. That we were able to identify high variability in a network believed to subserve social cognition provides an important proof of principle that S-GPFA is able to identify meaningful variance within a heterogeneous sample. We expect that S-GPFA will enable important tests of both exploratory and hypothesis-driven examinations of functional topography in psychiatric disorders.

% \vspace{-5pt}
\subsubsection{Group-specific temporal dynamics}
\begin{figure}[t]
    \centering
    \includegraphics[width=1\linewidth]{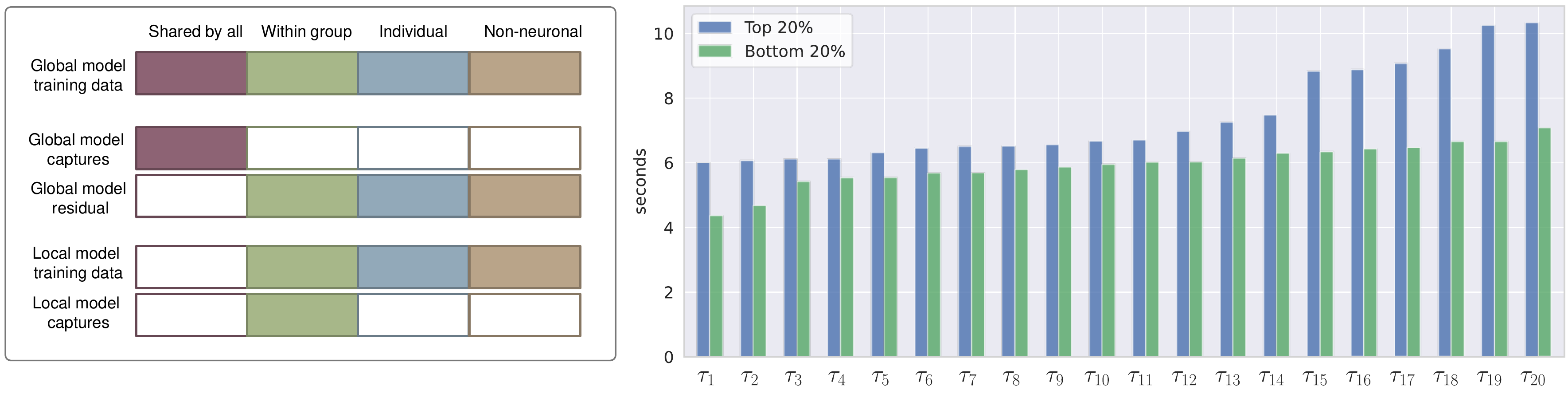}
    % \vspace{-0.5em}
    \caption{Group-specific temporal dynamics. Left: Different data components captured using the global and local models. Note that the local models operate on the residuals of the global model. Right: Group-specific timescales (sorted by magnitude) discovered using local models, after removing temporal patterns shared across all subjects.}\label{fig:gtd}
    % \vspace{-10pt}
\end{figure}

Next, we employed S-GPFA to examine group-specific temporal dynamics. Our method is illustrated in the left of Figure \ref{fig:gtd}. First a \textit{global} model is trained using subjects from all groups. Residuals of the global model, as defined by $\Y{m} - \w{m} \X$, for $m \in \{1, \ldots, \m \}$, allow us to remove components of the observations that are shared among all groups \cite{srm}. Next, separately on each group, two models are trained over the residuals of the global model. These \textit{local} models capture group-specific dynamical components in observations after the globally shared components of the data are removed. 

We applied this approach to the SPINS dataset, opting to split the sample into two smaller groups comprised of individuals demonstrating EA scores within the top and bottom 20th percentiles, respectively. This bifurcation allowed us to isolate subjects with distinctly strong and poor socioemotional cognition, irrespective of schizophrenia diagnosis \cite{insel2014nimh}. The right of Figure \ref{fig:gtd} shows group-specific timescales discovered via the local models, after extraction of the residuals from the global model, i.e., after all signal with a global structure has been removed. In the local models, we observe that the group with the strongest EA performance shows less temporal variability, i.e., the dynamical patterns in those individuals with the greatest socioemotional cognitive capacity unfolded over slower timescales, in contrast to those participants with comparatively impoverished capacity. This observation is consistent with broad evidence of a relationship between domain-specific task performance, and observed modularity versus flexibility of task-relevant brain networks, e.g.\cite{olsen2013functional}, and the recent dissociation that strong performance on tasks involving little executive function or cognitive control (consistent with EA task demands) may be subserved by modular network activation \cite{ramos2017static}. Importantly, the isolation of low-dimensional group-specific timescales allows for the identification of differences of small effect, which may be statistically occluded in global models. 
% \paragraph{Finding distinct time points of the task}
% We present another application of the proposed model as an exploratory data analysis tool. Shared latent trajectories $\{\X_{p, :}\}_{p=1}^\p$ learned in S-GPFA summarize the underlying common state of participants during the course of performing a task. Hence, finding distinct or divergent moments in data can be beneficial in analyzing the collaborative response of subjects to the stimuli. To this end, we can simply find the pairwise distance between every pair of two time points $t_1, t_2 \in \{1, \ldots, \t\}$ of latent trajectories $\mathcal{D}(\X_{:, t_1}, \X_{:, t_1})$ in order to discover distinct time points during the task. Figure \ref{fig:dist_time} shows an example of such exploratory analysis for EA videos in the SPINS dataset.  
% \begin{figure}[h!]
%     \centering
%     \includegraphics[width=0.5\linewidth]{figures/dist_time_9.pdf}
%     \vspace{-1em}
%     \caption{[?]}\label{fig:dist_time}
% \end{figure}

\section{Conclusion}
\label{sec:conclusion}

In this paper, we introduced S-GPFA, a novel application of Gaussian Process Factor Analysis for finding subject-specific topographies and shared temporal dynamics in multi-subject fMRI datasets. First, we delineated the need for amplifying the smoothness loss in the training objective, which was supported by evaluation on simulated data. Then, we conducted three sets of experiments on human fMRI datasets, designed in turn to show that the dynamical structures found by our model are (i) shared between all subjects; (ii) generalizable to new subjects; and (iii) may be extended to isolate dynamical elements within groups. The proposed model is, to the best of our knowledge, the only analytic model that simultaneously allows functional aggregation, dimensionality reduction, and dynamical modeling of fMRI data. 

This work is motivated by the growing interest in modelling dynamics in fMRI, which has increased alongside the recognition that functional topographies vary substantially across individuals. The main limitation of S-GPFA is scalability with respect to the number of time samples. In practice, it is possible to divide long recordings into smaller chunks, and feed them as multiple samples to a single model. In future work, we may further refine the method by adopting (approximate) Bayesian inference approaches to draw uncertainty of parameter estimates, and using change-point and non-stationary kernels. Nonetheless, S-GPFA is well-poised for immediate impact on cognitive and psychiatric imaging studies aiming to discover shared neural signatures.

\paragraph{Broader Impact}
\label{sec:imapact}
The search to identify neural structure and function underlying human cognition is the holy grail of cognitive neuroscience. In psychiatry, the impetus of this work is more than mere description: the discovery of structural or functional correlates predictive of impairment may prove the basis for a given disorder. Crucially, identification of such correlates introduces testable neural targets for pharmacological or behavioural intervention studies aimed to abate or reverse disease course: a critical step forward in improving quality of life for those resistant to existing treatment.

% \clearpage
\section*{Supplementary Materials}
\beginsupplement
\section{Notations}
\label{sec:notation}

Scalar variables are denoted by regular small letters ($a$), constants are drawn using regular capital letters ($A$), vectors are defined with bold small letters ($\bm{a}$), and matrices are denoted by bold capital letters ($\bm{A}$).
\begin{table}[h!]
  \caption{Notations used in paper}
  \label{table_suppmat:notations}
  \centering
  \begin{tabular}{ll}
    \toprule
    \cmidrule(r){1-2}
    Variable     & Description\\
    \midrule
    $\A_{i,:}$       & The $i$th row of a matrix $\A$. It is a row vector. \\
    $\A_{:,j}$       & The $j$th column of a matrix $\A$. It is a column vector. \\
    $\left[\A\right]_{i,j}$     & Element $(i, j)$ of a matrix $\A$. \\
    $\m$              & Number of subjects in the dataset. \\
    $\q$              & Number of brain regions/voxels in each brain image. \\
    $\t$              & Number of TRs for subjects. \\
    $\p$              & Number of latent dimensions. \\
    $m$             & Index for subjects, $m\in\{1, \ldots, \m\}$. \\
    $q$             & Index for brain regions/voxels, $q\in\{1, \ldots, \q\}$. \\
    $t$             & Index for TRs, $t\in\{1, \ldots, \t\}$. \\
    $p$             & Index for latent dimensions, $p\in\{1, \ldots, \p\}$ \\
    $\Y{m}$         & $\q$ dimensional fMRI time series of length $\t$ for subject $m$: $\Y{m}\in \reals^{\q\times\t}$. \\
    $\X$            & Shared latent time series for all subjects. $\X\in \reals^{\p \times \t}$. \\
    $\w{m}$         & Factor loading matrix/functional topographies for subject $m$.  $\w{m} \in \reals^{\q \times \p}$. \\
    $\sv{m}$        & Noise factor for subject $m$. \\
    $\rv{q}$        & Noise factor for region $q$. \\ 
    $\spcov{m}$     & Observation noise covariance across brain regions, for subject $m$: \\
                    & $\spcov{m} = \sv{m}\operatorname{diag}(\rv{1}, \ldots \rv{\q}) \in \reals^{\q \times \q}$. \\
    $\kappa_p$      & Mercer kernel function for the GP prior over $p$th latent trajectory. \\
                    & We used a Square Exponential Kernel in this work. \\
    $\K_p$          & Covariance matrix for the $p$th latent trajectory, across time dimension. $\K_p \in \reals^{\t \times \t}$. \\
    $\alpha_p^2$    & SE kernel variance for the $p$th latent trajectory.  \\
    $\eta_p^2$      & SE kernel independent noise variance for the $p$th latent trajectory. \\
    $\tau_p$        & SE kernel characteristic timescale for the $p$th latent trajectory. \\
    $\lambda$       & Smoothness loss amplifier. \\
    $\params$       & The collective set of model parameters and latent variables to be inferred: \\
                    & $\params=\left\{ \X, \{\w{m}, \sv{m}\}_{m=1}^M, \{\rv{q}\}_{q=1}^Q, \{\tau_p\}_{p=1}^P \right\}$.\\
    \bottomrule
  \end{tabular}
\end{table}

\section{Additional Simulated Dataset}

Here, we use sinusoidal latent trajectories to generate simulated data. We used $\p = 4$ sinusoidal series as latent trajectories with different frequencies. We also sampled factor loading weights from a standard normal distribution, independently for each of $M=20$ subjects. Linear combination of latent trajectories and factor loadings generates $Q = 30$ dimensional times series of length $T = 200$ for each subject. Furthermore, we set $\{\rho_{(m)}\}_{m=1}^\m$ and $\{\sigma_{q}\}_{q=1}^\q$ to linearly spaced values in interval $(0.1, 0.5)$.
Noisy versions of these observations are used to train S-GPFA and SRM models.

\begin{figure}[h!]
    \centering
    \includegraphics[width=\linewidth]{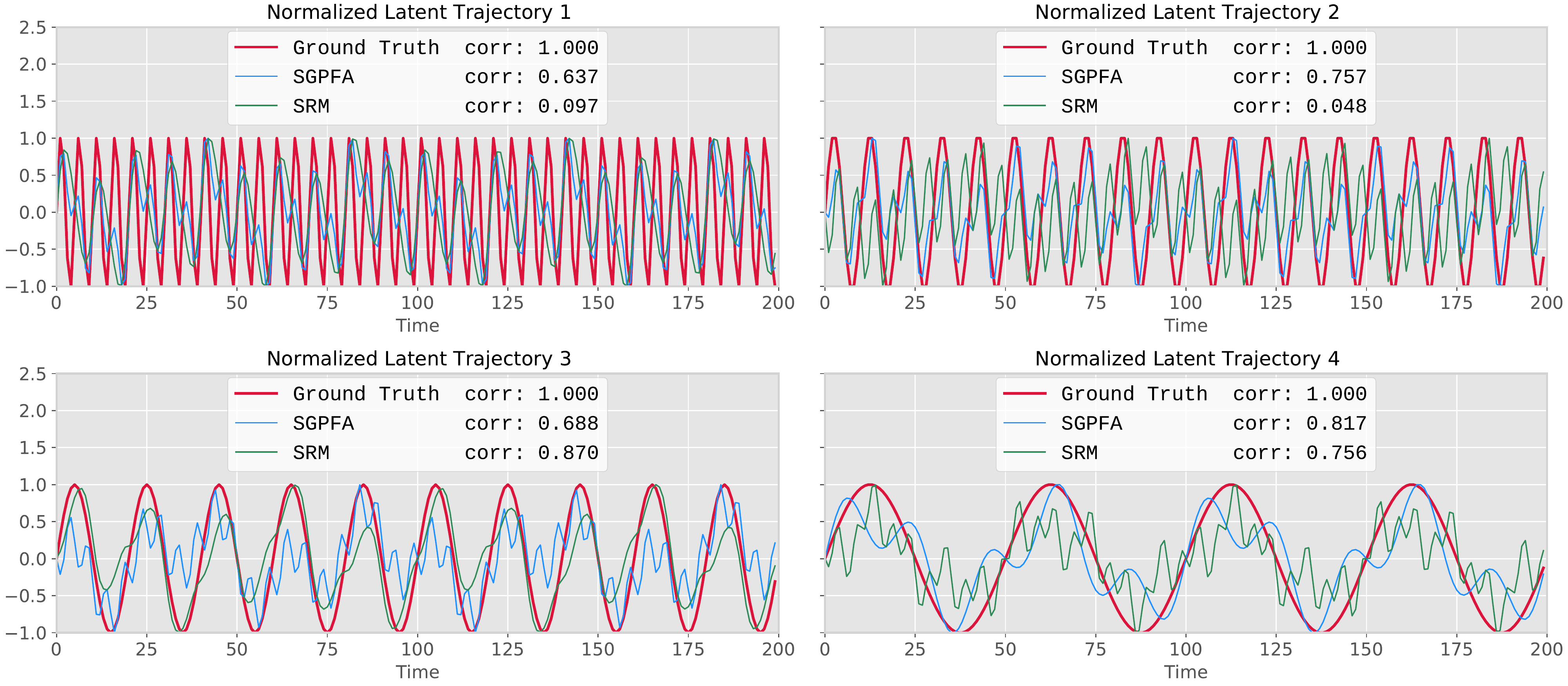}
    \includegraphics[width=\linewidth]{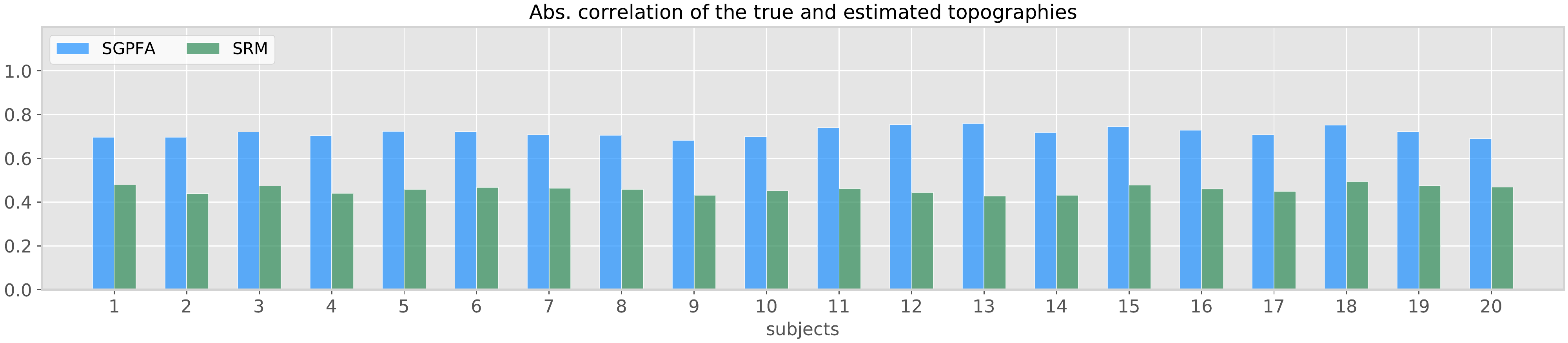}
    \caption{$\lambda=1$: Simulated dataset with parameters $M=20$, $Q=30$, $T=200$, $P=4$. sinusoidal functions with frequencies $[.02, .05, .1, .2]$ are used as latent trajectories to generate data.} \label{fig_suppmat:sim_sgpfa_1}
    \vspace{15pt}
    \includegraphics[width=\linewidth]{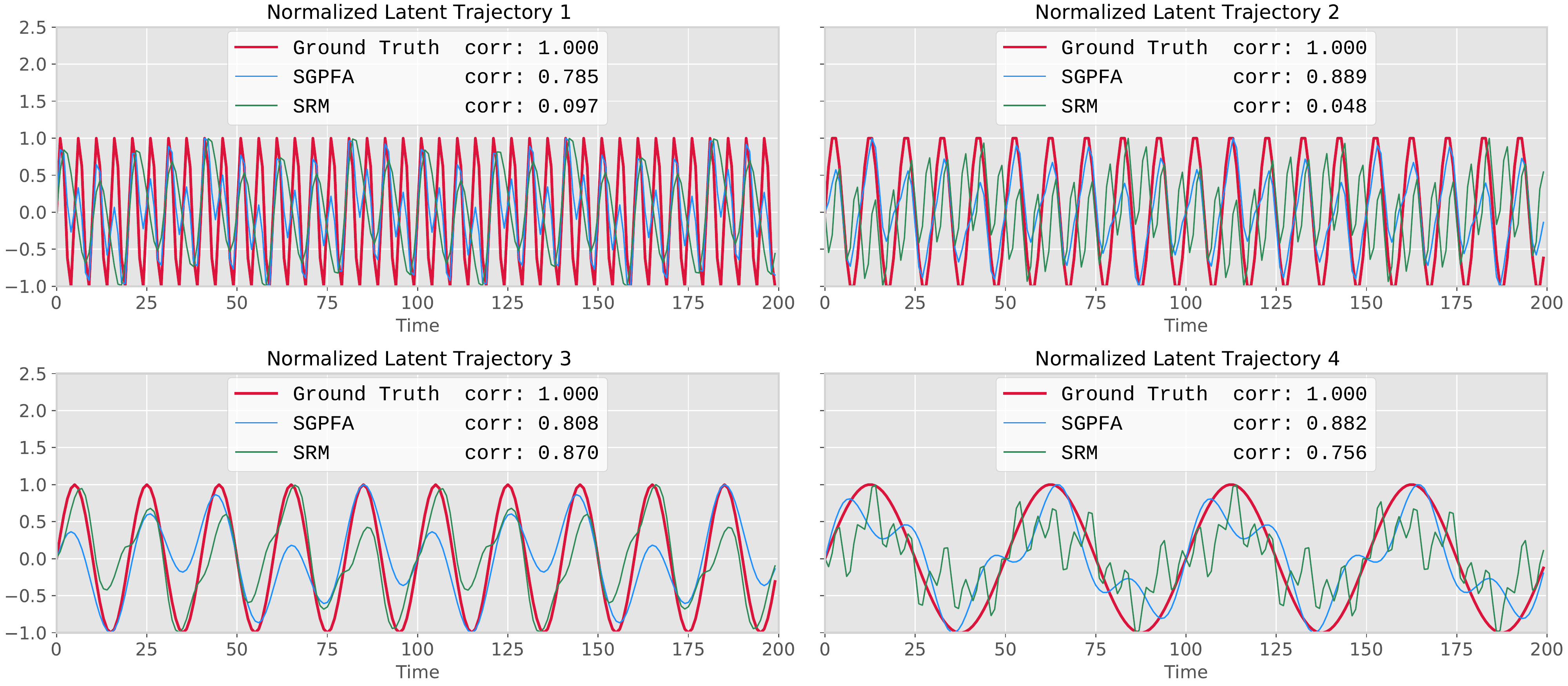}
    \includegraphics[width=\linewidth]{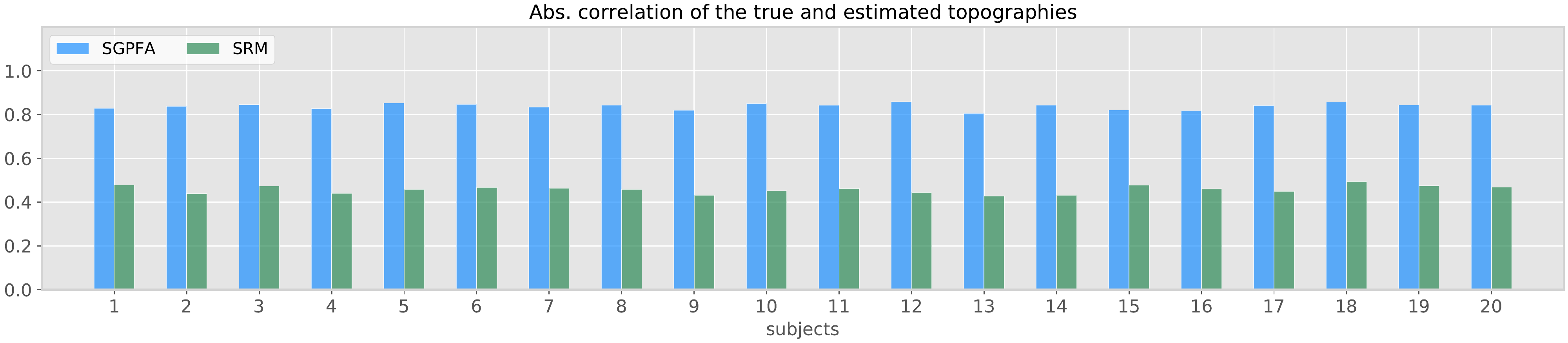}
    \caption{$\lambda=0.1 \times \m\q/\p=15$. Same dataset as in Figure \ref{fig_suppmat:sim_sgpfa_1}. Amplified smoothness loss results in finding more accurate parameters.} \label{fig_suppmat:sim_sgpfa_2}
\end{figure}

\section{fMRI Datasets}

Table \ref{table_suppmat:datasets} summarizes essential information about each dataset used in this paper.

\begin{table}[h!]
    \caption{Datasets used in the paper}
    \label{table_suppmat:datasets}
    \centering \small
    \begin{tabular}{lcccc}
        \toprule
        \cmidrule(r){1-2}
        Dataset & Subjects  & TRs   & Voxels/Regions    & ROI\\
        \midrule
        SPINS (EA Task) \cite{spins} & 332   & see Table \ref{table_suppmat:ea}  & 400 regions \cite{schaefer2018local} & whole brain \\
        Raider \cite{ha}  & 10    & 2203  & 1000   & ventral temporal cortex (VT) \\
        Sherlock \cite{sherlock} & 17  & 1976   & 481 & posterior medial cortex (PMC) \\
        \bottomrule
    \end{tabular}
\end{table}

\subsection{SPINS dataset}
We used a subset (332 subjects) of the NIMH Data Archive study “Social Processes Initiative in the Neurobiology of the Schizophrenia(s)” (SPINS) who completed the Emotional Accuracy (EA) task with usable recordings. We selected SPINS as an experimental dataset because it includes participants with and without schizophrenia, and we held that the anticipated variability in brain structure, function, and cognitive performance would provide an interesting test of S-GPFA.

\subsubsection{Acquisition and Preprocessing}
The fMRI acquisition was an echo-planar sequence, with TR=2000ms, TE=30ms, voxel size=3mm isotropic, 50 volumes, and a 64364 matrix; importantly, all parameters were matched as closely as possible across the study’s three centres and six scanners, within limitations of hardware. We removed the first 3 TRs of each of the 9 video runs, regressed known confounds, detrended mean values, and applied a low pass (0.08) and high pass (0.009) filter. Confound regression adjusted for six rigid-body motion parameters (i.e., transformation and rotation in each 3-dimensional plane), two nuisance confounds estimated by physiological fluctuations in white matter and cerebral spinal fluid, and the first three components based on anatomical noise correlation methods calculated by fMRIprep \cite{fmriprep}. We did not apply spatial smoothing. The anatomical image used in preprocessing was a fast-gradient echo T1 with repetition a 2300ms repetition time, 0.9 mm isotropic voxels, no slice gap, and an interleaved ascending acquisition. We parcellated the brain into 400 functionally-defined cortical regions of interest using the Schaefer atlas \cite{schaefer2018local}, interpreted in conjunction with the Yeo 17 networks \cite{yeo}.

The SPINS dataset had, at time of data request, 451 consented subjects that continued to meet eligibility criteria over three study visits. Of these, 403  subjects completed the 3 fMRI acquisitions required by the 9-video EA task, and 372 subjects provided usable EA ratings data (i.e. complied with the task, did not fall asleep). Of these, 332 subjects were found to have usable data (we discarded data gauged to be of insufficient quality on the basis of technician error, irreparable scanner artifacts, and/or high participant motion, defined as a mean framewise displacement of .5mm or greater over the course of any EA video). The final dataset of 332 subjects included 187 individuals with and 145 individuals without schizophrenia.

\subsubsection{EA Task}
The EA task collects fMRI as participants watch videos of an actor (‘target’) recounting autobiographical events. In total, 9 videos were shown, lasting for between 2-2.5 minutes. Participants provide ratings of the target’s valence on a 9-point scale (1=extremely negative, 9=extremely positive) in real time via button press. The tasks' primary dependent measure, the EA score, is the correlation between the participant’s ratings of the targets’ emotions, and the “gold standard” rating of the targets’ ratings of their own emotions, calculated in 2 second time epochs. Table \ref{table_suppmat:ea} presents information regarding each EA video.

% \begin{table}[h!]
%   \caption{EA videos}
%   \label{table_suppmat:ea}
%     \centering
%     \begin{tabular}{ c c c c }
%         \toprule
%         \cmidrule(r){1-2}
%         Valence & Emotion & Description & Timepoints \\ 
%         \midrule
%         negative & sad & soccer & 66 \\
%         positive & amused & movie & 70 \\
%         positive & delighted & trip & 91 \\
%         positive & amused & comedian & 74 \\
%         negative & anger & paycheeck & 74 \\
%         negative & sad & death & 83 \\
%         negative & anger &  truck & 56 \\
%         positive & delighted &  wedding & 73 \\
%         negative & anger & roommate & 61 \\
%         \bottomrule
%     \end{tabular}
% \end{table}

\begin{table}[h!]
  \caption{EA videos}
  \label{table_suppmat:ea}
    \centering
    \begin{tabular}{c l l l l c}
        \toprule
        \cmidrule(r){1-2}
        Video & Valence &	Emotion &	Description &	Narrator &	Timepoints \\ 
        \midrule
        1 & positive &	delighted &	trip &	female &	85 \\
        2 & negative &	sad &	soccer &	male &	73 \\
        3 & positive &	amused &	comedian &	female &	55 \\
        4 & negative &	anger &	paycheck &	male &	72 \\
        5 & positive &	delighted &	wedding &	male &	69 \\
        6 & positive &	amused &	movie &	male &	73 \\
        7 & negative &	sad &	death &	female &	59 \\
        8 & negative &	anger &	room-mate &	female &	89 \\
        9 & negative &	anger &	truck &	female &	64 \\
        \bottomrule
    \end{tabular}
\end{table}

\section{Consistency of Functional Topographies over 9 EA Videos}
\label{sec_suppmat:topo_consistency}
\begin{figure}[h!]
    \centering
    \includegraphics[width=1\linewidth]{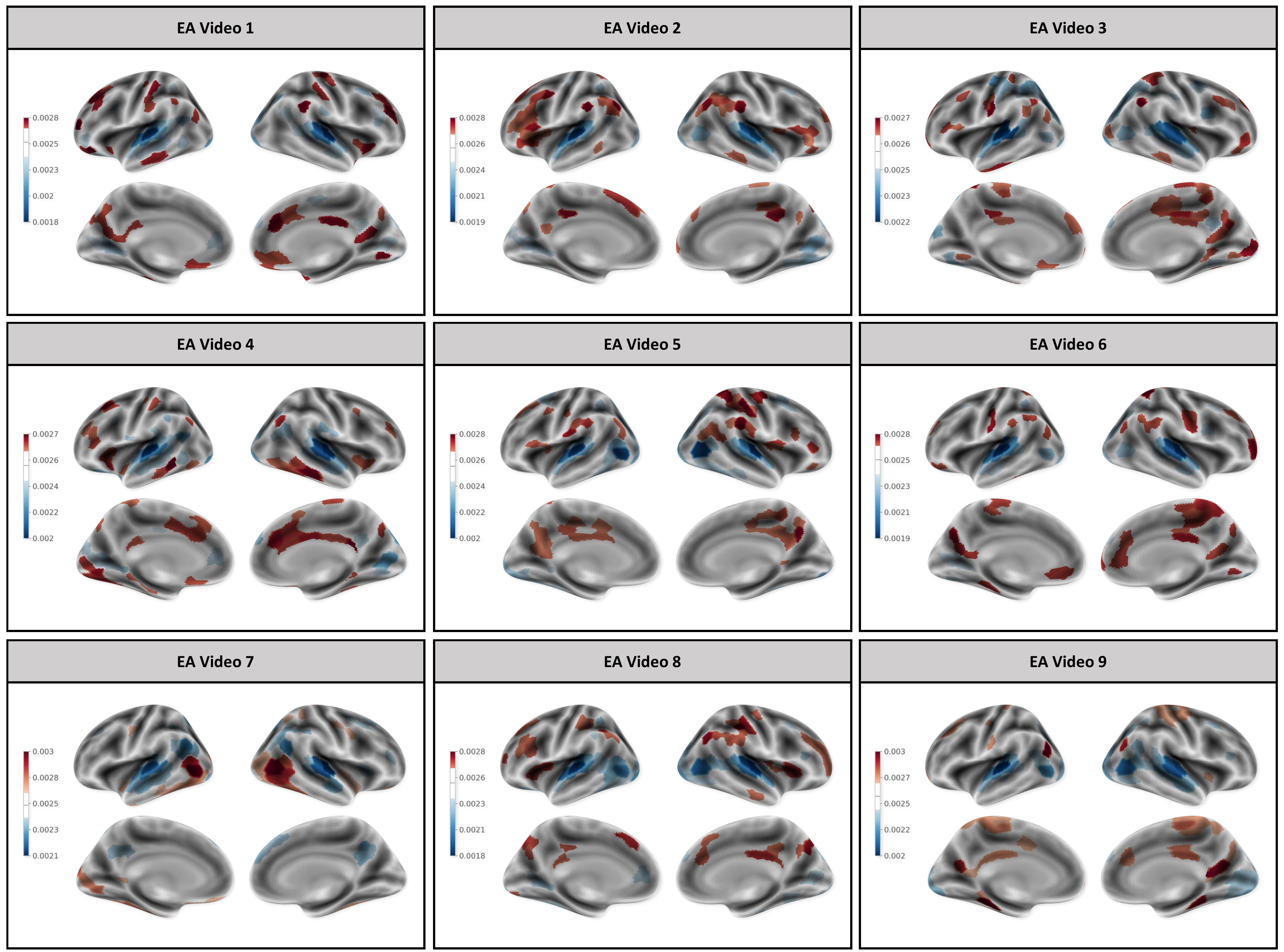}
    \caption{Consistency of functional topographies -- 9 EA Videos. Brain surface plots of average variance of normalized topographies, for left and right hemispheres. Blue (red) indicates higher (lower) consistency across subjects. }\label{fig_supmat:topo_all}
\end{figure}

Figure \ref{fig_supmat:topo_all} presents the result of the analysis in section \ref{subsec:topo_consistency} for all nine EA videos in the SPINS dataset.

\begin{figure}[h!]
    \centering
    \includegraphics[width=0.65\linewidth]{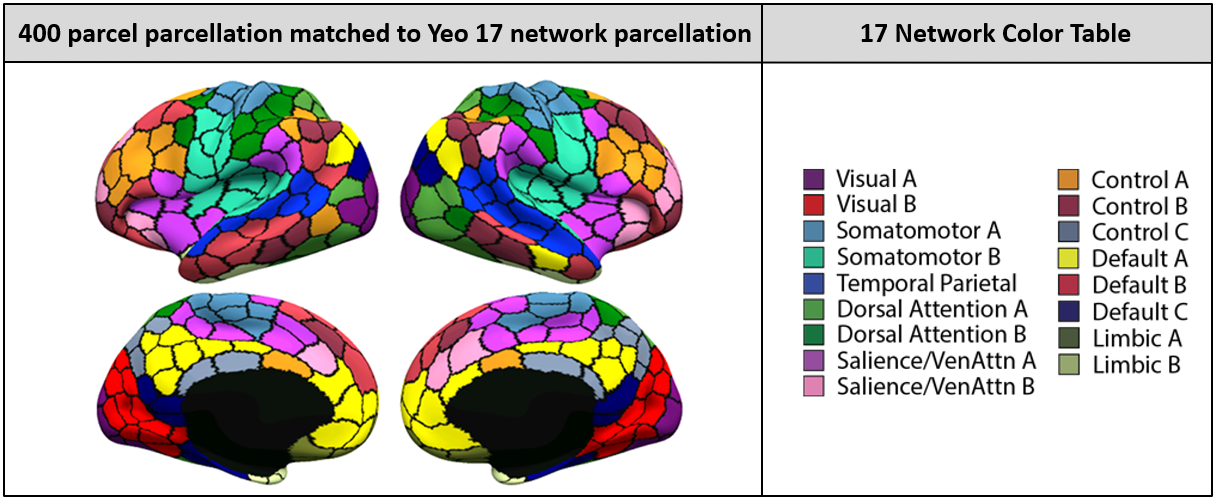}
    \caption{Visualization of the Schaefer 2018 Local-Global 400 parcellation in fslr32k space. Parcels were colored to match Yeo 17 network parcellation \cite{yeo}. Taken from \cite{schaefer2018local}.}\label{fig_supmat:schaefer}
\end{figure}

% \begin{ack}
% Use unnumbered first level headings for the acknowledgments. All acknowledgments
% \end{ack}

\clearpage
\bibliography{ref}

\end{document}